\documentclass{mem}
\usepackage{natbib}\usepackage{txfonts}\usepackage{balance}
\usepackage{graphicx}
\usepackage[a4paper]{hyperref}
\begin{document}
\def\teff{$T\rm_{eff }$}
\def\kms{$\mathrm {km s}^{-1}$}

\title{X--ray emission from Clusters of Galaxies}

% \subtitle{}

%\author{P. \,Tozzi\inst{1} }
\author{P. \,Tozzi}

\offprints{P. Tozzi}

\institute{Istituto Nazionale di Astrofisica -- Osservatorio
Astronomico di Trieste, Via Tiepolo 11, I-34143 Trieste, Italy --
\email{tozzi@oats.inaf.it} }

\authorrunning{Tozzi}

% \titlerunning{X--ray Clusters}

\abstract{In the last eight years, the Chandra and XMM--Newton
satellites changed significantly our view of X--ray clusters of
galaxies. In particular, several complex phenomena have been directly
observed: interactions between cluster galaxies and the Intra Cluster
Medium (ICM), cold fronts in the ICM, hot bubbles due to relativistic
jets from radio loud AGN, the lack of cold gas in ``cool--cores'', and
non--thermal X--ray emission.  Still, this increasing complexity does
not prevent us from using X--ray clusters as a tool to constrain
cosmological parameters.  In addition, observations of clusters up to
redshift $\sim 1.3$, allowed us to trace the thermodynamical and
chemical evolution of the ICM on a time interval as large as 8 Gyr.
In this presentation, I will give a personal introduction to the most
debated topics in this field, to end with some prospects for the
next--generation X--ray satellites.  \keywords{X--rays -- Galaxy:
Clusters -- Cosmology: observations } }

\maketitle{}
% \maketitle{X--ray emission from Clusters of Galaxies}

\section{From simplicity to complexity}

Clusters of galaxies are the largest virialized objects in the
Universe.  They form via gravitational instability from the initial
perturbations in the matter density field.  Clusters are made out of
three main ingredients: non--collisional dark matter ($\sim 80$\%),
hot (and warm) diffuse baryons ($\sim 17$\%), and cooled baryons
(stars in galaxies or diffuse stars, $\sim 3$\%).  The dark matter is
dynamically dominant and form potential wells in which the baryons are
trapped.  The collapse leads to violent relaxation: dark matter and
baryons rapidly adjust and reach a pressure balance with the
gravitational forces.  The velocities of particles inside the halos
become randomised, and all the matter components share the same
equilibrium within the virial radius.

The diffuse baryons are heated to temperatures between $T_X \sim 1$
keV and $T_X \sim 10$ keV (roughly corresponding to masses ranging from
$10^{14}$ to $5 \times 10^{15} M_\odot$) and constitute the Intra
Cluster Medium (from now on ICM).  Thanks to the very low density of
the electrons (typically $n_e \sim 10^{-3}$ cm$^{-3}$) the ICM is
optically thin and it is in a state of collisional equilibrium
established between the electrons and the heavy ions.  The resulting
emission from the ICM in the X--ray band is described by a continuum
component due to thermal Bremsstrahlung (roughly scaling as $\propto
T^{1/2} n_e^2$) plus line emission from K--shell and L--shell
transitions of heavy ions (Iron being the most prominent).  The
collisional equilibrium allows one to derive directly from the X--ray
spectrum of the ICM both the electron temperature and the abundances
of heavy elements \citep{k05}.

Since the discovery of the X--ray emission from clusters of galaxies
by the Uhuru satellite, X--ray cluster samples have been considered
one of the best tools for cosmology, thanks to the simple link between
the X--ray spectral properties and the total dynamical mass (as a
consequence of the virial theorem) and to the simple selection
function depending only on the flux limit.  Now that we are in the
maturity of the Chandra/XMM--Newton era, we realize that clusters
harbor an unexpected complexity, that demands a much deeper
understanding of the thermodynamics of the ICM and its relation with
the other mass components, such as member galaxies and dark matter.
This is a well known effect: as soon as you look at Nature with better
instruments, it reveals increasing complexity\footnote{In better
words: ''There are more things in heaven and earth, Horatio, Than are
dreamt of in your philosophy."}.

The present challenge is to update the models of the thermodynamics of
the ICM, including several complex processes, in order to re-establish
on a firmer basis the reputation of clusters as ``laboratories for
galaxy evolution'' and ``signposts for cosmology''.  In this
perspective, I will briefly review what I consider the most relevant
issues to address in order to better understand the ICM physics.
These are:

\begin{itemize}

\item interactions between galaxies and ICM;

\item the ``cool--cores'' problem;

\item non thermal X--ray emission;

\item chemistry and thermodynamics of the ICM at high redshifts.

\end{itemize}

Finally, I briefly discuss what we would like to do with the next
generation of X--ray satellites.

\begin{figure}[t!]
\centering \includegraphics[height=4.9cm]{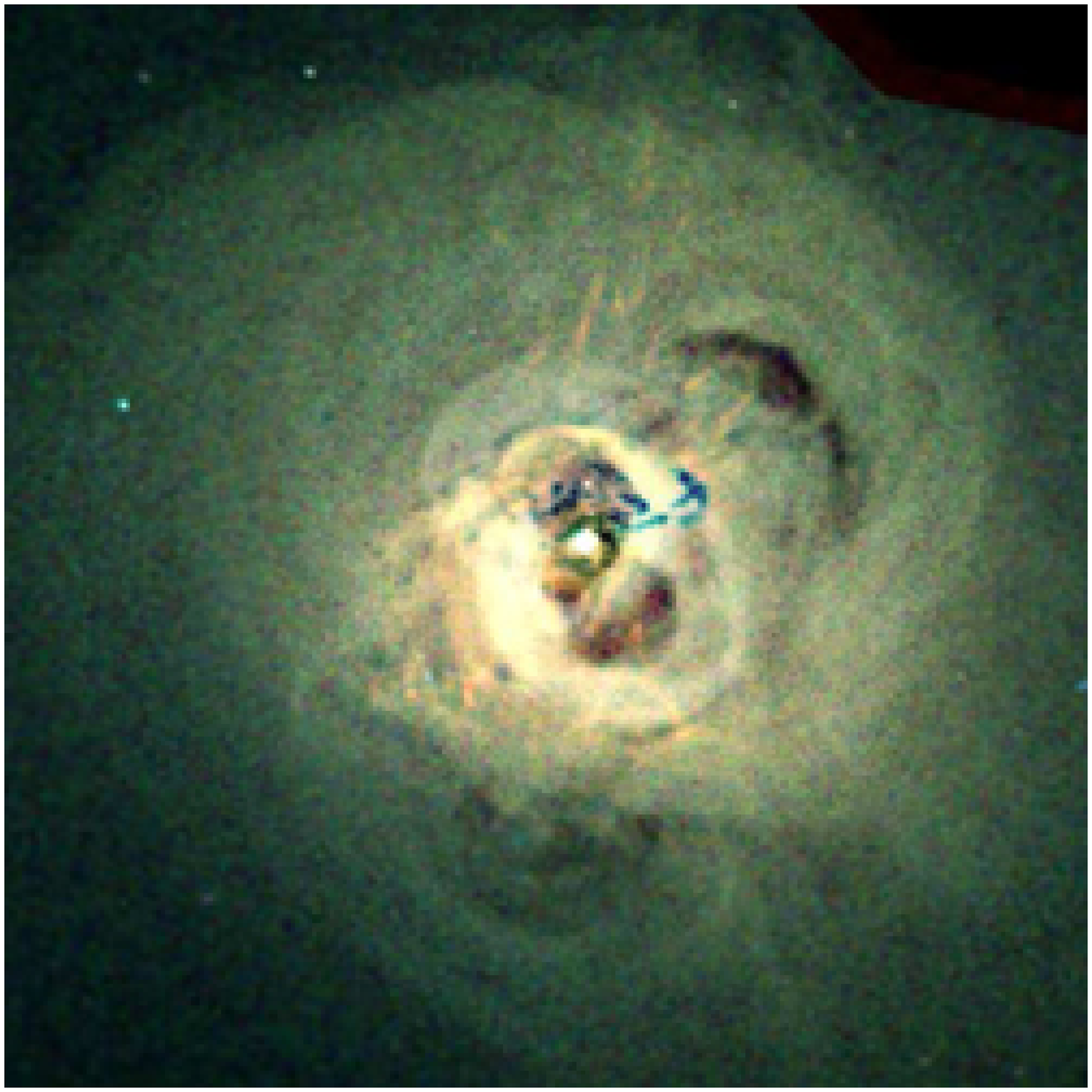}
\centering \includegraphics[height=5cm]{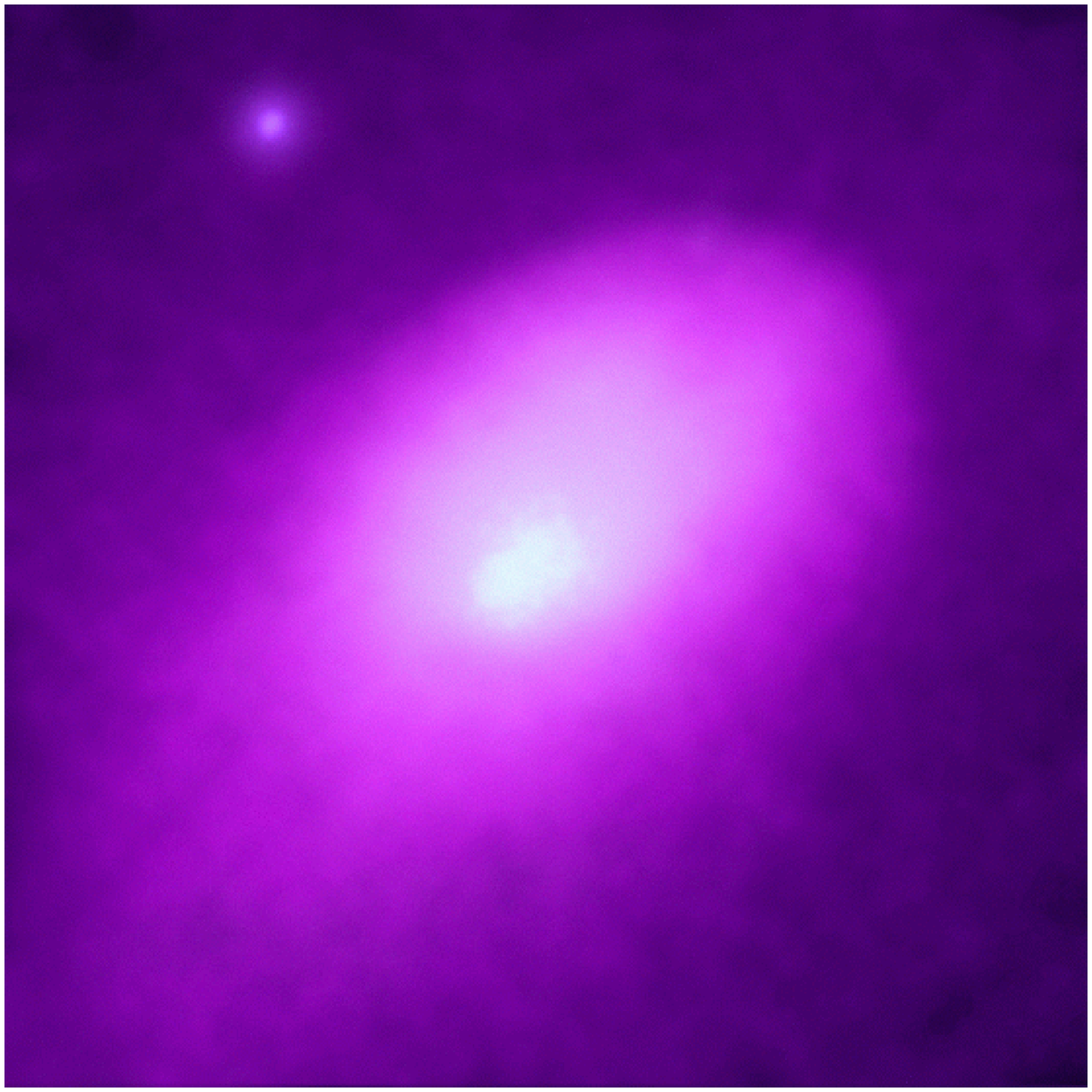}
\centering \includegraphics[height=5cm]{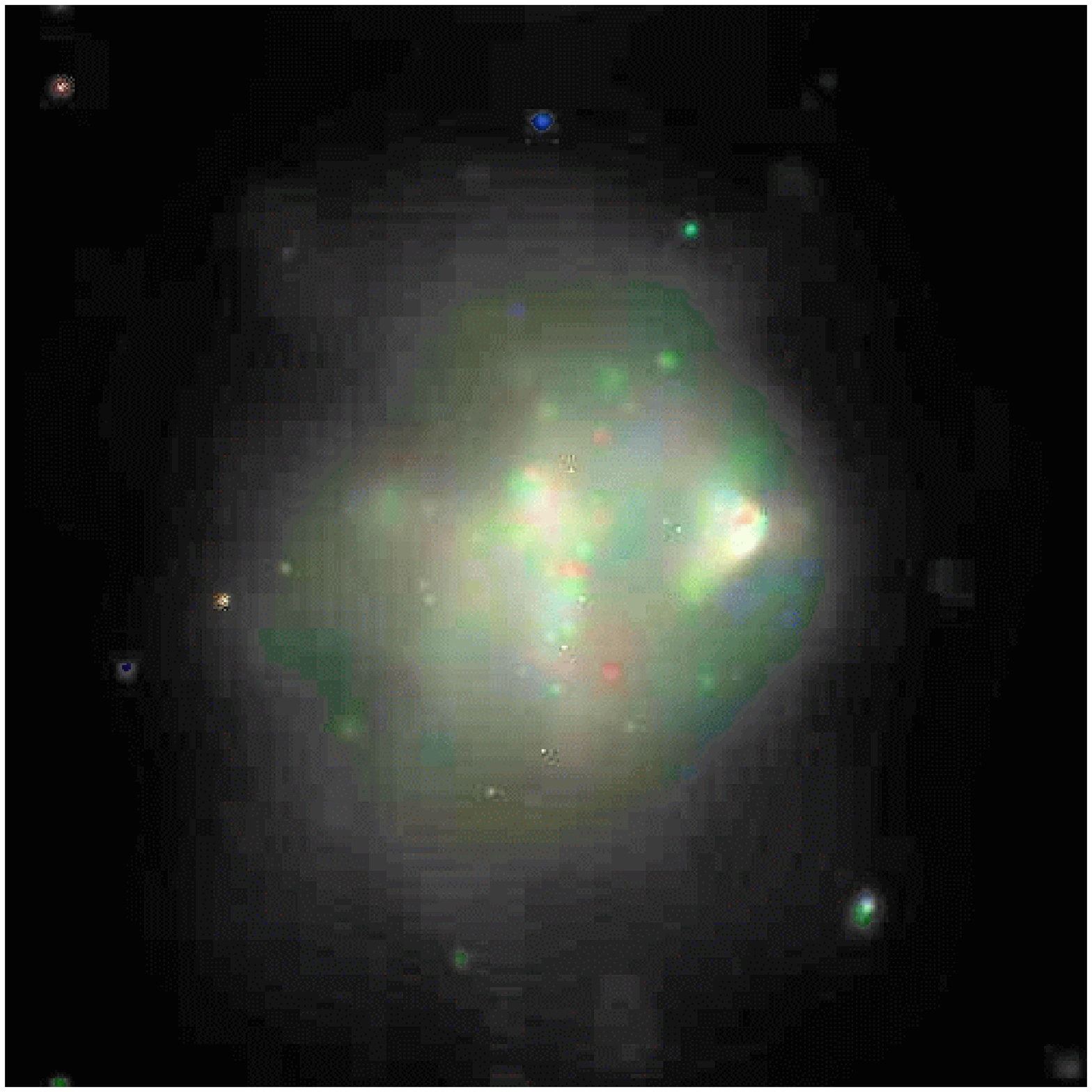}
% If not, use
%\picplace{5cm}{2cm} % Give the correct figure height and width in cm
\caption{\footnotesize A short gallery of Chandra images of X--ray
clusters with ongoing interactions between ICM, galaxies and dark
matter.  From top to bottom: cavities in the ICM of the Perseus
cluster due to the jets from the central radio loud AGN \citep{fab03};
cold fronts in Abell 2142 \citep{mark00}; an ongoing massive merger in
1E0657--56, the {\sl bullet cluster} \citep{mark04}.}
\label{bullet}
\end{figure}

\section{Interactions between ICM, Cluster Galaxies and dark matter}

As we said above, the naive picture of clusters of galaxies envisages
the baryons as a thermal plasma smoothly sitting in the dark matter
potential well, with the galaxies as test particles flying around
without any relevant effect on the surrounding medium.  We know now
that galaxies do affect the ICM, not only because of their motion, but
mostly because of energetic internal processes like star formation and
nuclear activity.

The most dramatic view of galaxy/ICM interactions is given by the
gosthly surface brightness distribution of the Perseus cluster (Fabian
et al. 2003), shown in the first panel of Figure \ref{bullet}.  The
radio loud AGN in the central galaxy is responsible for the two large
simmetric cavities, originated by the relativistic electrons of the
radio jet that pushed away the ICM.  Other cavities in the ICM at
larger distance from the center show that this episode is somehow
recurrent.  The mechanical energy associated to these events is huge,
and may have a dominant role in giving energy to the ICM in excess
with respect to that associated to virialization.  However, it is not
clear how this mechanical energy can be transformed into thermal
energy of the ICM.  The presence of an AGN in the central galaxy is
also expected to be relevant for the solution of the cool--cores
problem (see below), and more in general to raise the average entropy
level of the ICM with respect to the self--similar scaling
\citep{pon03}.

The second panel of Figure \ref{bullet} shows cold fronts visible as
sharp discontinuities in the surface brightness distribution of Abell
2142.  The presence of {\sl cold fronts} has been discovered in
Chandra images of bright nearby clusters \citep{mark00,maz01}.
Despite the jump in the electron density, the pressure gradient is
constant across the front.  The cold fronts are probably due to
subsonic motion of colder, group sized clumps of gas.  Maybe they are
the final stages of a major merger before the onset of a new
equilibrium configuration.  These observations warn against the use of
a simple beta model \citep{cff76} to describe the gas distribution and
to derive the total mass of a cluster, as indicated also by
significant differences between the X--ray and the lensing mass
determinations found in some systems.

In some cases clusters show clear signs of ongoing massive mergers
strongly affecting the dynamical equilibrium.  A spectacular example
is 1E0657--56, the {\sl bullet cluster} \, \citep{mark04}, shown in
the third panel of Figure \ref{bullet}.  Comet--like structures such
as the ``bullet'' are bow shocks associated to ultrasonic motions.  In
these non--equilibrium stages, the distribution of the diffuse,
pressure--supported baryons can significantly differ from that of the
dark matter, with significant implications on the nature of the dark
matter itself \citep{doug06}.

\section{The cool--cores enigma}

More than half of X--ray clusters of galaxies present central spikes
in the surface brightness distribution.  The temperature in these
central regions is significantly lower than the average one.  Due to
the higher electron density, the cooling time is significantly lower
than the dynamical time of the cluster.  Therefore, it is reasonable
to expect that the ICM is actually cooling at a rate directly related
to the X--ray luminosity, as in the case of the isobaric cooling
model.  One of the prediction of the so called ``cooling--flow''
model, is that a wide range of temperatures must be present in the
center of such clusters, down to the lowest temperatures detectable in
the X--ray band (few tenths of keV).  Due to the widespread presence
of heavy elements in the ICM, the low temperature gas cools mainly
through line emission.

\begin{figure}[t!]
\centering \includegraphics[height=5.7cm]{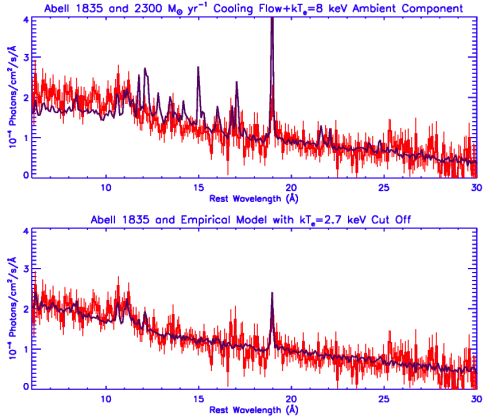}
% \centering \includegraphics[height=5.7cm]{coolcorenew.ps}
\caption{\footnotesize The top panel shows the model (blue) and the
data from the Reflection Grating Spectrometer of XMM--Newton for the
cool--core cluster Abell 1835.  In the bottom panel the gas colder
than 2.7 keV has been removed in the model, which now shows a good
agreement with the data \citep{pet06}.}
\label{cc}
\end{figure}

It was a big surprise when the analysis of the high resolution X--ray
spectra of the ``cooling flows'' (taken with the Reflection Grating
Spectrometer onboard XMM--Newton) strongly contradicted this picture.
It turned out that the emission lines associated to the coldest gas
were missing, as shown in Figure \ref{cc} \citep{pet06}.  The lack of
the emission lines implies the lack of cold gas.  It is possible to
estimate a lower limit to the temperature of the gas in the center of
X--ray clusters, which always turns out to be about a third of the
average (or virial) temperature.  Since the gas is cooling but only
down to a minimum temperature, these regions have been dubbed
``cool--cores'' as opposed to ``cooling--flows''.  Istantaneously the
nature of ``cool--cores'' became one of the most enigmatic issue in
the field of X--ray clusters.

Which is the process that keep the temperature of the ICM above $1/3
\, \, T_{vir}$?  The presence of some extra--energy in the ICM was
already known from the study of the scaling relations between X--ray
observables like temperature and luminosity.  Gravitational processes
(shock heating and virialization) predict a self--similar relation of
the kind $L_X \propto T^2$, as confirmed by hydrodynamical N--body
simulations.  The observed $L_X\propto T^3$ relation, as well as the
observation of an entropy excess in the ICM with respect to the
self--similar scaling \citep{pon99}, already demonstrated that the ICM
can be well described with an extra heating of about 1 keV per
particle in addition to the virial energy.

However, the modeling of non--gravitational heating is very difficult,
since the cooling is a runaway process which depends on the square of
the electron density, while many energetic process we may think of,
scale linearly with the total mass density.  We have two obvious
candidates which can inject energy associated to non--gravitational
processes: the prime candidate is feedback from star formation
processes, whose effects on the ICM are clearly shown by the presence
of heavy elements in the ICM.  The ICM is polluted with SNe ejecta (in
particular Iron, seen in the X-ray spectrum at 6.7-6.9 keV
rest--frame) in an amount consistent with being produced by the
massive ellitpicals observed within the virial radius \citep{mv88}.
However it is quite hard to predict the amount of energy dumped into
the ICM by SNe explosions, since a large part of it can be radiated
away, depending on the physical state of the insterstellar medium
around star forming regions.  N--body hydrodynamical simulations
however, seem to indicate that star formation alone is not sufficient
to raise the entropy of the ICM to the observed level \citep{bo05}.

The second, and most promising, candidate is recurrent feedback from
nuclear activity in the cluster galaxies, as directly observed in some
nearby bright objects (as in Perseus, shown in Figure \ref{bullet}).
Relativistic jets from radio loud AGN may have the right amount of
energy, but, once again, the mechanism of transfer of the mechanical
energy into thermal energy of the ICM, and the duty cycle of this
process, are still unclear.

The Occam's razor pushes us to think that the same process is
responsible for both the break of self--similarity of X--ray scaling
laws and the temperature floor in cool--cores.  Not only, our love for
simplicity suggests that also the processes regulating star formation
in galactic halos have the same nature.  It is well known, in fact,
that in Cold Dark matter model, the process of galaxy formation from
baryons cooling in dark matter potential wells would cause a ``cooling
catastrophe'' without some sort of feedback which re--heats most of
the baryons and stop star formation \citep{mu02}.  Once again, jets
from radio loud AGN and SNe explosions are considered the main agents
responsible for the self--regulation of star formation everywhere in
the Universe.

To summarize, the evidences of non--gravitational heating of the ICM
(feedback) are:

\begin{itemize}

\item observation of jets/ICM interactions (direct);
\item emission lines of heavy elements in the X--ray spectra of the
ICM (direct);
\item temperature floor in cool--cores (indirect);
\item breaking of self--similar scaling (indirect);
\item the absence of catastrophic overcooling (indirect).
\end{itemize}

It is worth mentioning that the presence of the non--gravitational
energy input does not hamper the use of clusters as a tool for
cosmology.  The relation between X--ray observables and the dynamical
mass (which is crucial to link the distribution of clusters to
cosmological parameters as briefly described below), is still valid,
and maybe it is even tighter once we can provide a comprehensive model
for the ICM thermodynamics.  A recent example is the use of the
parameter $T_X M_{gas}$ which appears to be tightly correlated to the
dynamical mass \citep{kr06}.  Therefore, my feeling is that this
increasing complexity is not an impediment, but, on the contrary, is
giving us a more powerful tool to investigate at the same time many
aspects of structure formation, from subgalactic to cluster scales.
The price to pay is, of course, a bigger effort in describing the ICM
physics.

\section{Non--thermal X--ray emission}

It is now recognized that non thermal components are relevant in the
ICM.  The clearest evidence of a relativistic component comes from the
radio band.  In particular, some clusters show a large scale radio
halo, or radio relics, which are produced by a relativistic population
of electrons emitting via synchrotron in the presence of a magnetic
field.  Such relativistic electrons may be associated to acceleration
mechanisms occurring at the shocks during major mergers, or during
diffuse turbulence.

The study of these components can strongly benefit by the combined
observation of the radio and the X--ray emission.  In fact, the
relativistic electrons are expected to produce an hard tail in the
X--ray spectrum of the ICM mainly due to inverse Compton.  Hard tails
have been observed in few nearby clusters \citep{ff07}, despite their
existence is still controversial.

Even if the energy content of non--thermal components is limited to
few percents, as commonly thought, its effects are relevant to
understand the ICM physics.  For example, the evaluation of the
magnetic field is crucial to assess the role of thermal conduction,
which, in turn, can affect the structure of cooling cores, where
strong temperature gradients are present.

\begin{figure}[t!]
\centering \includegraphics[height=5cm]{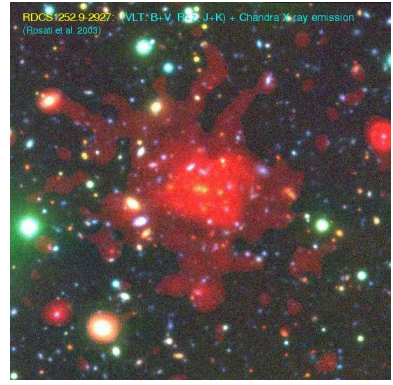}\\
\vskip 0.3cm
\centering \includegraphics[height=7.5cm]{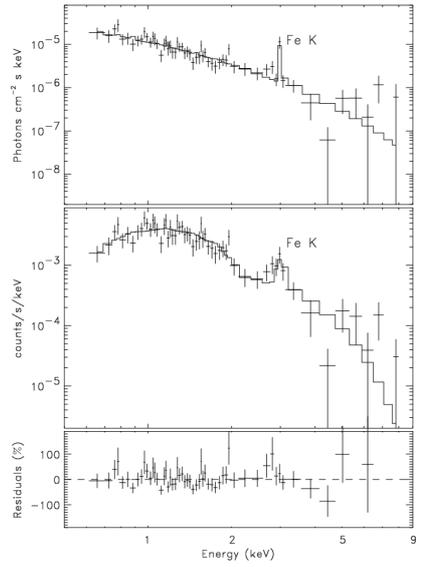}
\caption{\footnotesize The top panel shows the X--ray emission (in
red) of the cluster RXJ1252 at $z=1.235$ on top of the optical image.
Lower panels show the X--ray spectrum of RXJ1252, where the Iron
emission line is clearly visible \citep{ros04}.}
\label{rxj1252}
\end{figure}

\section{Clusters of galaxies up to $z \sim 1.3$}

Chandra and XMM--Newton deep observations allowed to study the ICM of
clusters at redshifts as high as $z\sim 1.3$.  One of the best example
is the Chandra observation of RXJ1252, where we clearly detected the
Iron line in the X--ray spectrum (see Figure \ref{rxj1252}).

\begin{figure}[t!]
\centering \includegraphics[height=6.2cm]{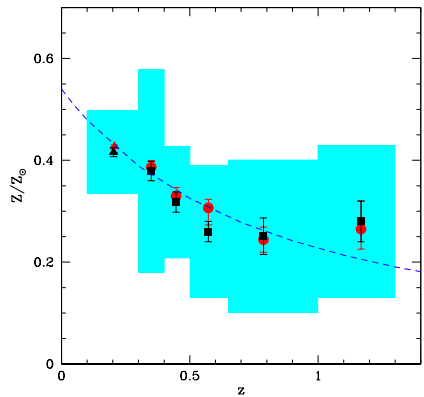}
\caption{\footnotesize Solid squares and circles show Iron abundance
in the inner regions of hot clusters ($kT>5$ keV) averaged over about
10 clusters in each bin (with two different combining methods), as a
function of redshift \citep{bal07}.  The shaded area show the rms
dispersion of the single measures, while the dashed line is
a fit to the $Z_{Fe}$ evolution of the form $\sim(1+z)^{-1.25}$.}
\label{bal}
\end{figure}

The presence of heavy elements in the ICM traces the distribution of
SNe products into the diffuse hot baryons.  Therefore, the evolution
of heavy elements abundances with cosmic epoch is a crucial
information concerning the interaction of the ICM with cluster
galaxies.  Precise measurements of the heavy elements content of
clusters over large look--back times provide a useful fossil record
for the past star formation history of cluster baryons.  So far, only
the abundance of Iron can be traced back to the highest redshifts
where X--ray clusters are observed.  Recently, Balestra et
al. presented results from a sample of 56 clusters at redshifts
$z>0.4$ observed by {\em Chandra} and XMM-{\em Newton} \cite{bal07}.
The average Iron abundance is already significant at $z\sim 1.3$, at a
look-back time of $\sim 9 $ Gyr, in line with the peak in star
formation for proto--cluster regions occurring at redshift $z > 2$.
In addition, an increase of the average Iron abundance with cosmic
time below $z=0.5$ has been observed in the central regions (see
Figure \ref{bal}).  The evolution in $Z_{Fe}$ with $z$ can be
parametrized by a power law of the form $\sim(1+z)^{-1.25}$.  The
observed evolution implies that the average Iron content in the inner
regions (radii $R \leq 0.2 R_{vir}$) of the ICM at the present epoch
is a factor of $\sim2$ larger than at $z\simeq 1.3$.  These data
provide significant constraints on the time scales of the physical
processes that drive the chemical enrichment of the ICM.  For example,
this evolution can be explained by the release of enriched gas into
the ICM from disk galaxies transforming into S0 in the central regions
of clusters \citep{cal07}.  However, the mechanism responsible for the
gradual transfer of the enriched gas from the cold, low entropy phase
associated to galaxy or group--size halos to the ICM is still under
investigation with hydrodynamical simulations (Cora et al. in
preparation).

% We find a significantly higher average iron abundance in
% clusters with $kT<5$~keV, in agreement with trends measured in local
% samples. For $kT>3$~keV, $Z_{Fe}$ scales with temperature as
% $Z_{Fe}(T)\simeq0.88\,T^{-0.47}$.

\section{Cosmology with X--ray Clusters}

\begin{figure}[t!]
\centering \includegraphics[height=4.8cm]{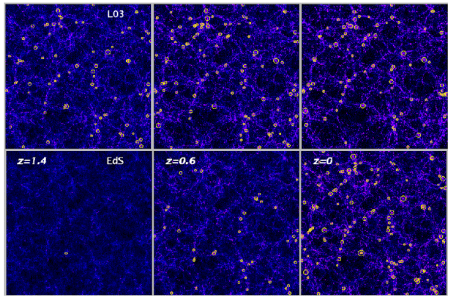}
\caption{\footnotesize A visual rendition of the hierarchical growth
of clusters of galaxies in a $\Lambda$CDM (first row) and a critical
(second row) Universe.  Massive clusters, indicated with small
circles, are abundant in a $\Lambda$CDM Universe even at redshift
$z\sim 1.4$.}
\label{bg01}
\end{figure}

Clusters are ideal objects to trace the large scale structure of the
Universe.  In particular, their number density as a function of the
mass scale and of the cosmic epoch strongly depend on cosmological
parameters.  The observation of clusters over a wide range of
redshifts is therefore a valuable tool for cosmology.  In Figure
\ref{bg01} we show the hierarchical evolution of hot ($kT>3$ keV)
clusters in different cosmologies, as it appears in N--body
simulations.  Given the same local abundance of clusters, in a flat
$\Lambda$CDM Universe (first row) there are several hot massive
clusters at redshift $z\sim 1.4$, as opposed to an high density
Universe (second row).  About fifteen years ago, before results from
deep surveys with the ROSAT satellite had been published \citep{r98},
the presence of massive, hot clusters at redshift as low as $z\sim
0.2$ was strongly questioned.  Now, the presence of massive clusters
at redshift as large as $z\sim 1.3$ is firmly established, and this
represents a strong hint in favour of a low density, $\Omega_0\sim
0.3$, $\Lambda \sim 0.7$ CDM Universe.  This is in line with what has
been found in several CMB experiments in recent years.

The distribution of clusters can be used to constrain cosmological
parameters thanks to well established models for the mass function of
virialized halos.  The simple theory of linear evolution of the
initial density-perturbations field plus the virial theorem applied to
the non--linear overdensities, allows one to compute straight out the
mass distribution of virialized halos as a function of cosmic epoch.
The Press \& Schechter theory \citep{ps} and its extensions, have been
widely tested with numerical simulations, and are now commonly used to
predict the temperature and luminosity distribution of X--ray
clusters.  The main uncertainty is now due to the relation between the
dynamical mass and the X--ray observables.  As discussed above, this
relation is not simple, but it is well established on an observational
basis and on theoretical grounds.  Quite clearly, the challenge now is
to refine the modeling of the mass/X--ray observables relations and
their intrinsic scatter, in order to perform the so called ``precision
cosmology'' with clusters.

\begin{figure}[t!]
\centering \includegraphics[height=3.2cm]{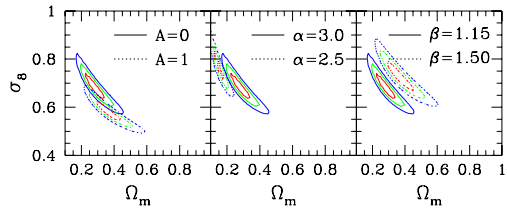}
\caption{\footnotesize Confidence levels in the $\sigma_8$--$\Omega_0$
plane from the X--ray luminosity function of distant clusters of
galaxies observed with ROSAT \citep{bor99}.  Different panels
correspond to different shapes of the linear power spectrum of density
fluctuations.}
\label{sigma8}
\end{figure}

However, X--ray clusters always proved to be an excellent tool for
cosmology.  An example of their effectiveness is given by the
constraints on the cosmological density parameter $\Omega_0$ and on
the normalization of the linear spectrum of density fluctuations
$\sigma_8$, derived using ROSAT data, as shown in Figure \ref{sigma8}.
In particular, we would like to remark that the value of $\sigma_8\sim
0.7$ found for clusters, initially at variance with the measure
$\sigma_8 = 0.84 \pm 0.04$ in the first year of WMAP, is now in
agreement with the new WMAP results \citep{sp07}.  Once again, this is
a good example of the complementarity of X--ray clusters with respect
to other geometrical cosmological tests, like high--z SNe and CMB
experiments.

\section{Great expectations}

Next future X--ray clusters astronomy must face a double challenge:
first, understanding the physics of the ICM, exploring in greater
details the cool--cores but also the low surface brightness regions in
the outskirts of clusters; second, finding more clusters and groups at
high redshifts.  It is a matter of fact that the large majority of
high--z X--ray clusters, which are the targets of present--day Chandra
and XMM--Newton deeper observations, have been discovered by ROSAT,
while only and handful have been discovered in recent times.

In order to find more high--z clusters, how much wide and deep do we
need to go?  A step forward will be provided by missions like eRosita
(PI G. Hasinger), which will provide a medium--deep, all--sky survey
with medium spatial resolution.  Another strategy is going deeper in a
smaller solid angle with a wide field imager with good spatial
resolution (few arcsec).  This is one of the goal of EDGE, a medium
size class mission, devoted to the study of diffuse baryons both in
emission or in absorption against bright GRB (PI L. Piro).  Deep
pointings with EDGE of about 1 Ms will reach flux limits comparable to
Chandra deep surveys but on a much larger solid angle.  EDGE will
allow one to detect groups out to $z>1$ and clusters out to $z\geq 3$.
In addition, another instrument onboard of EDGE, the Wide Field
Spectrometer based on cryogenic microcalorimeters, with excellent
energy resolution of about 3 eV will be able to provide direct
velocity measurements in the cluster cores to study turbulence in the
ICM and the cool--cores problem, as well as the faint outer regions of
clusters.

\section{Conclusions}

The physics of the ICM is complex, and it must include the combined
effects of star formation and nuclear activity in the cluster galaxies
on the diffuse baryons.  Understanding this complex interplay requires
a comprehensive approach to galaxy formation, mass accretion processes
onto Super Massive Black Holes, heavy elements production and
diffusion, AGN jets--ICM Interactions, effects of the non--thermal
components, and other phenomena, all at the same time.  The good news
is that understanding the ICM means also understanding galaxy
formation, and being able to constrain cosmological parameters and the
growth of the large scale structures across cosmic epochs.

To capitalize and extend what we have learned so far with Chandra and
XMM--Newton, we must have soon both a wide area medium--deep survey as
well as a mission devoted to the properties of the ICM.  Looking
further into the future, the next generation of X-ray telescopes must
achieve a spatial resolution comparable to that of Chandra, which
proved to be crucial to avoid confusion limit and to isolate genuine
diffuse emission at very high redshifts.

\begin{acknowledgements} 
We acknowledge financial contribution from contract ASI--INAF
I/023/05/0 and from the PD51 INFN grant.
\end{acknowledgements}

\bibliographystyle{aa}

\begin{thebibliography}{}

\bibitem[{Balestra et al. (2007)}]{bal07} Balestra, I., et al. 2007,
A\&A

\bibitem[{Borgani et al. (1999)}]{bor99} Borgani, S., et al. 1999, ApJ,
517, 40

\bibitem[{Borgani \& Guzzo (2001)}]{bg01} Borgani, S., \& Guzzo,
L. 2001, Nature, 409, 39

\bibitem[{see Borgani et al. (2005)}]{bo05}Borgani, S., et al. 2005,
MNRAS, 361, 233

\bibitem[{Calura et al. (2007)}]{cal07}Calura, F., Matteucci, F., \& Tozzi,
P. 2007, 2007, MNRAS in press, astro-ph/0702714

\bibitem[{Cavaliere \& Fusco Femiano (1976)}]{cff76}Cavaliere, A., \&
Fusco-Femiano, R. 1976, A\&A, 49, 137

\bibitem[{Douglas et al. (2006)}]{doug06}Douglas, C., et al. 2006, ApJ,
648L, 109

\bibitem[{Fabian et al. (2003)}]{fab03}Fabian, A.C., et al. 2003,
MNRAS, 344L, 43

\bibitem[{see Fusco Femiano et al. (2007)}]{ff07}Fusco-Femiano, R.,
Landi, R., Orlandini, M. 2007, ApJ, 654L, 9

\bibitem[{see Kahn (2005)}]{k05}Kahn, S.M. 2005, in {\sl High--Energy
Spectroscopic Astrophysics}, Saas--Fee Advanced Course 30, Springer

\bibitem[{Kravtsov et al. (2006)}]{kr06}Kravtsov, A.V., Vikhlinin, A.,
Nagai, D. 2006, ApJ, 650, 128

\bibitem[{Markevitch et al. (2000)}]{mark00}Markevitch, M., et al. 2000,
ApJ, 541, 542

\bibitem[{Markevitch et al. (2004)}]{mark04}Markevitch, M., et al. 2004,
ApJ, 606, 819

\bibitem[{Matteucci \& Vettolani (1988)}]{mv88}Matteucci, F., \&
Vettolani, G. 1988, A\&A, 202, 21

\bibitem[{Mazzotta et al. (2001)}]{maz01}Mazzotta, P., et al. 2001,
ApJ, 555, 205

\bibitem[{see Muanwong et al. (2002)}]{mu02}Muanwong et al. 2002, MNRAS,
336, 527

\bibitem[{Peterson \& Fabian (2006)}]{pet06}Peterson, J.R., \& Fabian,
A.C. 2006, Phys.Rept., 427, 1

\bibitem[{Ponman et al. (1999)}]{pon99}Ponman, T. J., Cannon, D. B., \&
Navarro, J.F. 1999, Nature, 397, 135

\bibitem[{Ponman et al. (2003)}]{pon03}Ponman, T.J., Sanderson, A.J.R.,
Finoguenov, A. 2003, MNRAS, 343, 331

\bibitem[{Press \& Schechter (1974)}]{ps} Press, W. H., \& Schechter, P.
1974, ApJ, 187, 425

\bibitem[{Rosati et al. (1998)}]{r98}Rosati, P., et al. 1998, ApJL, 492,
21

\bibitem[{Rosati et al. (2004)}]{ros04}Rosati, P., et al. 2004, AJ, 127,
230

\bibitem[{Spergel et al. (2007)}]{sp07}Spergel, D.N., et al.  2007,
ApJS, 170, 377

\end{thebibliography}

\end{document}